\documentclass[12pt,twoside]{article}
\usepackage{correl}

\def\TheTitle{Stochastic Series Expansion Methods}
\def\TheHeading{Stochastic Series Expansion}
\def\TheAuthor{Anders W. Sandvik}
\def\TheInstitute{Department of Physics, Boston University, Boston, Massachusetts, USA \newline
Institute of Physics, Chinese Academy of Sciences, Beijing, China}
\def\TheAddress{\vskip70mm Chapter 16 of {\it Many-Body Methods for Real Materials} Edited by Eva Pavarini, Erik Koch, and Shiwei Zhang
Schriften des Forschungszentrums J\"ulich, Modeling and Simulation, Volume 9, ISBN 978-3-95806-400-3} 

\def\TheChapter{16}           

\begin{document}  
\MakeTitle           

The Stochastic Series Expansion (SSE) technique is a quantum Monte Carlo (QMC) method that is especially efficient for many quantum spin systems
and boson models. It was the first generic method free from the discretization errors affecting previous path integral based approaches.
These lecture notes will serve only as a brief overview of the SSE method, and readers who would like to implement it for some specific models are
recommended to consult some of the cited references for more details. 

\vskip2mm
In the introductory section, the representation of quantum statistical mechanics by the power series expansion of ${\rm e}^{-\beta H}$ will be compared
with the likely more familiar path integrals in discrete and continuous imaginary time. Extensions of the SSE approach to ground state projection and quantum
annealing in imaginary time will also be briefly discussed. The later sections will introduce efficient sampling schemes (loop and cluster updates)
that have been developed for many classes of models. A summary of generic forms of estimators for important observables will be given. Some recent applications
will also be reviewed.

\section{Overview of stochastic series expansion methods}

\subsection{Path integrals and series expansions}
\label{sec:sse1}

\index{path integral}
The most obvious way to construct a Monte Carlo (MC) simulation scheme for a lattice Hamiltonian $H$ is to start from the path integral
formulation of quantum statistical mechanics, where ${\rm e}^{-\beta H}$ is written as a product of imaginary-time evolution operators ${\rm e}^{-\Delta_\tau H}$
with a small ``time slice'', $\Delta_\tau = \beta/L$, for some large number $L$  of slices\cite{Feynman53,Suzuki76}. The partition function
$Z={\rm Tr} \{ {\rm e}^{-\beta H} \}$
can then be written as
\begin{equation}
 Z_{\rm PI} = \sum_{\{ \alpha \}} \langle \alpha_0| {\rm e}^{-\Delta_\tau H}|\alpha_{L-1}\rangle \cdots 
 \langle \alpha_2| {\rm e}^{-\Delta_\tau H}|\alpha_1\rangle \langle \alpha_{1}| {\rm e}^{-\Delta_\tau H}|\alpha_0\rangle,
\label{zpath}
\end{equation}
where the sum is over $L$ complete sets of states in a suitably chosen basis. Because the time step $\Delta_\tau$ of the slices is small,
the matrix elements can be computed
to some approximation with a total error of order $\Delta_\tau ^p$, where $p$ depends on the approximation used, e.g., $p=2$ with the lowest-order split-operator
(Suzuki-Trotter) decomposition of the exponential operators, ${\rm e}^{-\Delta_\tau (H_A+H_B)} \approx {\rm e}^{-\Delta_\tau H_A} {\rm e}^{-\Delta_\tau H_B}$
(for generic non-commuting terms $H_A$ and $H_B$). Several variants of such path-integral based QMC methods, often called {\it world line} (WL)
methods, were developed in the 1970s and 1980s \cite{Suzuki77,Hirsch82}.

\vskip2mm
There was also an earlier proposal by Handscomb, dating back to the early 1960s \cite{Handscomb62}, to instead start from the power series expansion of
${\rm e}^{-\beta H}$. Handscomb made the observation that the Heisenberg exchange interaction between two $S=1/2$ spins, $H_{ij} ={\bf S}_i \cdot {\bf S}_j$,
is a permutation operator, and that traces of strings of these operators for different spin pairs can be easily evaluated analytically. In the case of a
ferromagnetic coupling (i.e., with a negative sign in front of ${\bf S}_i \cdot {\bf S}_j$), the minus sign in the exponential is canceled. The powers of
the Hamiltonian can be further decomposed into strings of two-body (bond) operators, $H_{b_n} \cdots H_{b_2}H_{b_1}$, with all combinations of the operators
$H_b = H_{i(b),j(b)}$, $b \in \{1,\ldots,N_b\}$ on a finite lattice in any number of space dimensions $D$ (where $N_b=DL$ for nearest-neighbor interactions
on a $D$-dimensional simple cubic lattice). Denoting a string of $n$ operator indices $b_1,b_2,\ldots, b_n$ by $S_n$, the partition function is then
\begin{equation}
Z_{\rm H} = \sum_{n=0}^\infty \frac{\beta^n}{n!}  \sum_{S_n} {\rm Tr} \{ H_{b_n} \cdots H_{b_2}H_{b_1} \},
\label{zhand}
\end{equation}
where the traces of the strings of permutation operators are positive definite.
With a simple scheme to evaluate the traces in terms of permutation cycles, the operator strings can be sampled using a Metropolis MC method and various
thermodynamic expectation values can be evaluated \cite{Handscomb62,Lyklema82}. Among these expectation values, the total internal energy is particularly
important as it takes the simple form $E=\langle H\rangle = -\langle n\rangle/\beta$, which shows that the mean number of operators $n$ in the expansion
during the sampling process (which includes updates increasing and decreasing the string length $n$) scales as $\beta N$, $N$ being the total number of spins.
Thus, the expansion is convergent (as is also guaranteed since $H$ on a finite lattice has a bounded energy spectrum) and there is in practice a
``fuzzy bound'' on the expansion order $n$, proportional to $\beta N$, that will never be exceeded during the sampling process. Both the computer
storage requirement and the processor time for completing a full MC updating sweep then also scale as $\beta N$.

\vskip2mm
It is illustrative to compare Handscomb's expansion with the simple power series for the exponential of a positive number $x$,
${\rm e}^x = \sum_m x^m/m!=\sum_m W(m)$, which is always convergent and where the mean of the distribution $P(m)=W(m){\rm e}^{-x}$
(i.e., the Poisson distribution) is $\langle m\rangle = x$.
In light of this fact, the distribution of expansion orders in the case of ${\rm Tr}\{{\rm e}^{-\beta H}\}$ can be easily understood in the limit $\beta \to \infty$,
where ${\rm e}^{-\beta E_0}$ ($E_0$ being the ground state energy) is the only surviving contribution to the trace and $\beta |E_0|$ therefore corresponds
to $x$ above; $\langle n\rangle = \beta |E_0|$. At any temperature, the fluctuations of $n$ are related to the heat capacity;
$C=\langle n^2\rangle -\langle n\rangle ^2-\langle n\rangle$ \cite{Handscomb62}. Therefore, the variance of the distribution at $T=0$ is also exactly
the same as for the Poisson distribution; $\langle n^2\rangle -\langle n\rangle ^2 = \langle n\rangle$.

\vskip2mm
Handscomb's method is certainly elegant, but in its original formulation it was very limited in applicability, as there is only a small number of
models for which the traces can be computed analytically \cite{Chakravarty82}. The expansion is also normally not positive definite. The problem of
mixed positive and negative terms (i.e., an incarnation of the QMC sign problem) appears already for the antiferromagnetic Heisenberg interaction, and
it took some time until it was realized that this sign problem was easily solvable for bipartite lattices by simply adding a suitable constant to the
interaction \cite{Lee84}. The traces can then still be computed and the strings can be sampled in a way similar to the original formulation. However, in practice
the modified method did not perform as well as the path-integral based methods that had been developed by then for the Heisenberg antiferromagnet and many
other models. Though some further applications were reported \cite{Manousakis89}, Handscomb's method was essentially abandoned, as it became clear that it
was not possible in general to compute traces of operator strings efficiently, and, in the cases where the traces can be evaluated, the existing sampling
schemes were also often inefficient.

\vskip2mm
The dismissal of the idea of starting from the series expansion was premature. It is not clear why it was not realized earlier that it is not necessary
to compute the traces analytically---they also can be sampled in a chosen basis along with the operator strings \cite{Sandvik91,Sandvik92}. The sampling
weight in the extended space comprising states and operator strings has matrix elements $\langle \alpha |H_{b_n} \cdots H_{b_2}H_{b_1}|\alpha \rangle$ in
place of the full traces in Eq.~(\ref{zhand});
\begin{equation}
Z_{\rm SSE} = \sum_{n=0}^\infty \frac{\beta^n}{n!}  \sum_{S_n} \sum_\alpha \langle \alpha|H_{b_n} \cdots H_{b_2}H_{b_1}|\alpha \rangle.
\label{zsse1}
\end{equation}
Here the Hamiltonian has been defined as $H=-\sum_b H_b$, so that no negative signs appear explicitly. Of course the string of operators can still
produce negative signs, and the class of models for which this can be avoided is, as always, limited, but includes many important systems worthy of study
(some recent examples will be discussed in Sec.~\ref{sec:examples}). It can be noted here that sign problems originating from the diagonal part of the interaction
can always be avoided by adding suitable constants to some of the $H_b$ operators. Signs appearing with off-diagonal operations (either explicitly from negative
prefactors of some $H_b$ or due to fermion anti-commutation) are in general difficult to avoid \cite{Henelius2000}, except in special cases,
e.g., the aforementioned bipartite antiferromagnets, where the number of minus signs in the product is always even.

\vskip2mm
Methods based on sampling the traces in Eq.~(\ref{zsse1}) were first developed for general-$S$ Heisenberg models \cite{Sandvik91} and 1D Hubbard-type
models \cite{Sandvik92} almost 30 years after the advent of Handscomb's original method, and the extended series scheme eventually became known as the
SSE. Over the years, many further improvements of these algorithms have been made---some inspired by developments within other methods and some proposed
first within the SSE scheme and later adopted into other techniques. The SSE method was the first broadly applicable exact QMC scheme (i.e., not affected
by any systematical errors such as those from time discretization) and it was also the forerunner and inspiration to later algorithms based on other
series expansions, in particular, the perturbation series leading to the continuous-time {\it worm algorithm} \cite{Prokofev98}.

\vskip2mm
The SSE method can in principle be used for any model written in a discrete basis, though in practice sign problems restrict applications
to the same classes of models as the WL methods. Models that have been successfully studied include many spin Hamiltonians, boson models,
and 1D fermion systems (see Sec.~\ref{sec:examples}). Both the WL and SSE approaches normally have insurmountable sign problems for fermions in higher
dimensions (unlike, in some cases, auxiliary-field fermion determinant methods \cite{Blankenbecler81,Hirch85}). Models with controllable sign problems
accessible to SSE include certain frustrated impurities in antiferromagnets, where the signs arise only locally at the impurity and the bulk operators
do not involve any signs \cite{Liu09}. 

\vskip2mm
The sampled basis in SSE simulations is typically the $z$ components $S^z_i$ for spin systems or the site occupation
numbers $n_i=a^\dagger_ia_i$ for particle models. More complicated states can also be used, e.g., the basis of singlets and triplets on spin
dimers has been used to solve certain sign problems \cite{Alet16,Honecker16}. The primary advantage of SSE over discrete-time WL methods for sign-free models
is the absence of time-discretization error---the discrete dimension corresponding to the location in the SSE operator string
constitutes a faithful representation of continuous
imaginary time \cite{Sandvik92,Sandvik97}, as we will discuss below. Compared to more recently developed continuous-time WL methods \cite{Beard96,Prokofev98},
the discreteness of the SSE representation often allows for more efficient sampling schemes. For many models (especially spin systems), the SSE outperforms
all other known QMC methods. Typically the SSE method is also easier to implement.

\subsection{Continuous time in the power-series representation}

Before discussing how to sample the SSE configuration space (in Sec.~\ref{sec:sampling}) and compute expectation values of interest
(in Sec.~\ref{sec:estimators}), it is useful to consider some formal relationships between path integrals and the series representation
of statistical mechanics. At the same time we will introduce some concepts that will be useful later in algorithm definitions and implementations.

\vskip2mm
We can insert complete sets of states between all the operators in Eq.~(\ref{zsse1}) to write the SSE partition function in the form
\begin{equation}
 Z_{\rm SSE} = \sum_{n=0}^\infty \frac{\beta^n}{n!} \sum_{S_n} \sum_{\{ \alpha \}} \langle \alpha_0| H_{b_n}|\alpha_{n-1}\rangle \cdots
\langle \alpha_2|H_{b_2}|\alpha_1\rangle \langle \alpha_{1}|H_{b_1}|\alpha_0\rangle ,
\label{zsse2}
\end{equation}
where one can see a similarity with the path integral in Eq.~(\ref{zpath}). The inserted states (except for $|\alpha_0\rangle$) are redundant in the SSE
method, however, because the operators $H_b$ should be defined such that, for every possible combination of operator $H_b$ and basis state $|\alpha\rangle$,
the state $H_b |\alpha\rangle$  is proportional to a single basis state;
$H_b |\alpha\rangle = h_{b\alpha}|\alpha'\rangle$, $|\alpha'\rangle \in \{ |\alpha\rangle \}$
(and $h_{b\alpha}$ is the trivial matrix element $\langle \alpha '|H_b |\alpha\rangle$). Then the operator string itself uniquely defines how the state
$|\alpha_0\rangle$ in Eq.~(\ref{zsse2}) is propagated through a series of basis states (similar to a path in the WL formulation) and eventually arrives back
to the same state $|\alpha_0\rangle$ for an operator string contributing to the partition function (with the periodicity reflecting the original trace operation).
Clearly the vast majority of the strings violate the periodicity condition and for those we have $\langle \alpha_0| H_{b_n}\cdots H_{b_1}|\alpha_0\rangle =0$.
The strings should therefore be sampled so that the periodic ``time'' boundary condition is always maintained in any attempted update.

\vskip2mm
To ensure the ``no-branching'' condition $H_b |\alpha\rangle \propto |\alpha'\rangle$, the diagonal and off-diagonal parts of all the terms in $H$ have
to be separated. The index $b$ then does not just refer to sites or bonds (or larger units of sites for interactions involving more than two sites) but
enumerates separately the diagonal and off-diagonal terms. This is some times accomplished (formally and in actual program implementations) by
introducing an additional index, so that $H_{1,b}$ refers to, say, the diagonal part and $H_{2,b}$ is the off-diagonal part of the interaction on bond
$b$. In some cases the off-diagonal operators have to be formally split up further , e.g., in the case of $S>1/2$ Heisenberg spin systems the off-diagonal
part of the interaction, $S^x_iS^x_j + S^y_iS^y_j = \frac{1}{2} (S^+_iS^-_j + S^-_iS^+_j)$ has to be regarded as two different operators, $\frac{1}{2}S^+_iS^-_j$
and $\frac{1}{2}S^-_iS^+_j$. In contrast, for $S=1/2$ the sum of the two terms can be considered as a single operator, since only one
of them can give a non-zero result when acting on a basis state.

\vskip2mm
With the no-branching condition ensured, for a given operator string $S_n$, a propagated state $|\alpha(p)\rangle$ is defined in the SSE method
as the state obtained when the first $p$ operators have acted on the ket state $|\alpha_0\rangle = |\alpha(0)\rangle$ in Eq.~(\ref{zsse2}),
\begin{equation}
|\alpha(p)\rangle = r\prod_{i=1}^p H_{b_i} |\alpha_0\rangle,
\label{pstate}
\end{equation}
where $r$ formally is a normalization constant (which does not have to be computed in practice). In the sum over complete sets of states in
Eq.~(\ref{zsse2}), there is now a single contributing component for each $p$, namely, $|\alpha_p\rangle = |\alpha(p)\rangle$, and the operator
string uniquely defines the path of states evolving from $|\alpha_0\rangle$.

\vskip2mm
The propagated states also have a simple relationship to imaginary-time evolution in path integrals, where starting from some
state $|\phi (0)\rangle$ at imaginary time $\tau=0$ we have the evolved state $|\phi (\tau)\rangle = r{\rm e}^{-\tau H}|\phi(0)\rangle$, where
again we formally have a normalization constant $r$ because of the non-unitary evolution. For an SSE string of length $n$, we roughly have the correspondence
$(p/n)\beta \approx \tau$ (which becomes increasingly accurate with increasing $n$).

\vskip2mm
The precise relationship between the discrete index $p$ and the continuous time variable $\tau$ can be seen in the SSE expression for a time dependent
correlation function of some operators $A$ and $B$. With the a time dependent operator given by $B(\tau)={\rm e}^{\tau H} B {\rm e}^{-\tau H}$, by series
expanding the thermal expectation value $\langle B(\tau)A(0)\rangle$ we obtain 
\begin{equation}
\langle B(\tau)A(0)\rangle = \frac{1}{Z} \sum_{\alpha}\sum_{m,p}\frac{(\tau-\beta)^m(-\tau)^p}{m!p!} \langle \alpha| H^m B H^p A |\alpha \rangle,
\label{atau}
\end{equation}
which can also be expanded into a summation over operator strings as we did in the partition function above. We defer until later the question of what
types of operators can be considered in correlation functions in practice (but note that diagonal operators can always be treated easily). Here we just
consider the general formal relationship between the index $p$ in Eq.~(\ref{atau}) and the time difference time $\tau$. We can see that the expansion
of ${\rm e}^{-\tau H}$ in powers $H^p$ is independent of the expansion of ${\rm e}^{-(\beta-\tau) H}$ in  powers $H^m$, and to estimate the distribution
$P(p)$ roughly we can just replace $H^p$ by $\langle H\rangle^p$. Then we again have a Poisson distribution with mean $\langle p\rangle = \tau |\langle H\rangle|$,
which also equals $(\tau/\beta)\langle n\rangle$, where $n$ is the total expansion power of a given term; $n=m+p$. Stated differently, propagation by $p$
operators in SSE corresponds roughly to a displacement $\tau = (p/n)\beta$ in imaginary time, as already mentioned at the  end of the preceding paragraph.
Eq.~(\ref{atau}) shows more precisely that the relationship between any given $\tau$ involves summation over a sharply peaked distribution of states
propagated by powers $H^p$ of the Hamiltonian.

\vskip2mm
The similarity between the series approach and the path integral becomes even more obvious if the exponentials in Eq.~(\ref{zpath}) are expanded to
linear order, ${\rm e}^{-\Delta_\tau H} \approx 1-\Delta_\tau H$;
\begin{equation}
 Z_{\rm PI} \approx \sum_{\{ \alpha \}} \langle \alpha_0|1-\Delta_\tau H|\alpha_{L-1}\rangle \cdots 
 \langle \alpha_2|1-\Delta_\tau H|\alpha_1\rangle \langle \alpha_{1}|1-\Delta_\tau H|\alpha_0\rangle,
\label{zpath2}
\end{equation}
where the total error arising from all slices is of order $\Delta_\tau$, worse than the $\Delta_\tau^2$ error when the Trotter decomposition is used.
We will in the end take the limit $\Delta\tau \to 0$ and the treatment becomes exact. Next we again write $H$ as a sum over its individual terms $H_b$
with a minus sign taken out in front, $H=-\sum_b H_b$. Furthermore, we can introduce new name for the unit operators $1$ in the matrix elements, defining
$H_0=1$. Then, similar to the SSE formalism, we can write the path integral as a sum over index sequences $S_L = b_L,\ldots,b_2,b_{1}$ but now with the
new index $b=0$ also allowed and with the number $L$ of indices fixed;
\begin{equation}
Z_{\rm PI} = \sum_{\{ \alpha\}} \sum_{S_L}\Delta_\tau^n \langle \alpha_0|H_{b_L}|\alpha_{L-1}\rangle \cdots 
\langle \alpha_2|H_{b_2}|\alpha_1\rangle \langle \alpha_{1}|H_{b_1}|\alpha_0\rangle,
\label{zpath3}
\end{equation}
where $n$ is the number of elements in $S_L$ that are not equal to $0$ (i.e., the number of times actual terms of $H$ appear in a given operator
product).

\vskip2mm
In the case of the SSE partition function (\ref{zsse2}), there is an explicit sum over expansion orders $n$ that does not appear in the path integral
expression (\ref{zpath3}), but since we know that the series expansion is convergent we can introduce a cut-off, $n_{\rm max} = M$, and for expansion orders
lower than $M$ we can use ``fill-in'' unit operators, defining $H_0=1$ as above. If we further consider all possible ways of distributing $n$ Hamiltonian
terms within a product of $M$ operators out of which $M-n$ are unit operators, the SSE partition function becomes
\begin{equation}
 Z_{\rm SSE} = \sum_{\{ \alpha \}}  \sum_{S_M} \frac{\beta^n(M-n)!}{M!} \langle \alpha_0| H_{b_M}|\alpha_{M-1}\rangle \cdots
\langle \alpha_2|H_{b_2}|\alpha_1\rangle \langle \alpha_{1}|H_{b_1}|\alpha_0\rangle ,
\label{zsse3}
\end{equation}
where we have divided the weight by in Eq.~(\ref{zsse2}) by the combinatorial factor $M!/[n!(M-n)!]$ to compensate for overcounting of identical
contributions with different locations of the unit operators. Note again that $n$ is the number of non-$0$ elements in the operator-index string, and
a summation over $n$ in Eq.~(\ref{zsse3}) is implicit in the summation over all fixed-length sequences.

\vskip2mm
For a given common string length $M=L$, we now see that the path integral and SSE partition functions in Eqs.~(\ref{zpath3}) and (\ref{zsse3}) involve exactly
the same configuration space, but the weighting of the configurations is slightly different. However, the weights become identical in the limit where
$\Delta_\tau \to 0$ is taken in the path integral,
since $\Delta^n = \beta^n/L^n$ and for $M \to \infty$ we have $\beta^n(M-n)!/M! \to \beta^n/M^{n}$ in the SSE weight. Thus, we conclude that the two approaches
are really identical if the limit $M=L \to \infty$ is taken. An important difference is that the SSE approach is in practice exact (i.e., the truncation error is
exponentially small and not detectable in practice) already for some $M$ of order $\beta N$, while the path integral, in the approximation used above, is
affected by an error of order $\beta /L$. An exceedingly large number of slices $L$ would have to be used for the error to become completely negligible.
In a sense, the SSE method automatically finds the optimal number of slices for given $N$ and $\beta$.

\vskip2mm
Of course, the path integral approach as described above should not be used in an actual WL algorithm, because simple approximants of ${\rm e}^{-\Delta_\tau H}$
with smaller errors are available, i.e., the Trotter decomposition. The reason for using the linear approximation here was simply to obtain the most direct
relationship between the SSE and path-integral forms of $Z$. In the case of the SSE, while it is not necessary to introduce the fill-in operators
$H_0=1$ \cite{Sandvik92}, in practice it is actually convenient and efficient to use this device to achieve a fixed length of the index sequence. The
cut-off $M$ can be easily adjusted automatically during the equilibration part of a simulation, in such a way that the number $n$ of Hamiltonian operators
always stays below $M$ by a safe margin, as will be further discussed when updating schemes are described in Sec.~\ref{sec:sampling}.

\vskip2mm
The more recently developed continuous-time WL methods can be formally obtained by taking the limit $\Delta_\tau \to 0$
in Eq.~(\ref{zpath}). This is equivalent to the form (\ref{zpath3}) of the partition function, where the events correspond to the operators $H_{b_i}$.
However, in this case it is better to keep the diagonal operators in the exponential form instead of expanding them to linear order, and then the
paths of events dictated by the off-diagonal operators formally correspond to the perturbation expansion in the interaction representation
\cite{Prokofev98,Sandvik97}. In a computer program, only the ``events'', where and how the paths change, need to be stored \cite{Beard96,Prokofev98}.

\vskip2mm
To see the equivalence with the perturbation series more clearly, we can divide the $H$ into its diagonal part $H_0$ in the chosen
basis and an off-diagonal part $V$. As an example, for a generic Heisenberg spin system we could choose $H_0=\sum_{ij} J_{ij} S^z_iS^z_j$
and $V=\sum_{ij} J_{ij} (S^+_iS^-_j  + S^-_iS^+_j)$. Then, by considering $V$ as a perturbation to $H_0$ (though eventually there will be no restriction
on the strengths of the two terms) and carrying out the perturbation series to all orders we obtain the partition function as an integral in continuous
imaginary time
\begin{equation}
Z_{\rm CT} = \sum_{n=0}^\infty (-1)^n \int_0^\beta d\tau_1 \int_0^{\tau_1} d\tau_2 \cdots  \int_0^{\tau_{n-1}} d\tau_n
{\rm Tr}\{e^{-\beta H_0} V(\tau_n)  \cdots V(\tau_2)V(\tau_1)\},    
\end{equation}
where the time-evolved operator in the interaction representation is $V(\tau)={\rm e}^{\tau H_0} V{\rm e}^{-\tau H_0}$. Like in the SSE method,
we can now sample the trace in the chosen basis and write the product $V(\tau_n) \cdots V(\tau_2) V(\tau_1)$ as a string of time evolved
operators $V_b$ (now only including off-diagonal terms). When inserting complete sets of states between all the operators and summing over
diagonal terms for the trace, the exponentials just become numbers, and the weight of a given configuration (a state $|\alpha_0\rangle$ acted on
with an operator string) for a given set of times $\tau_1,\ldots,\tau_n$ can be easily computed. The integrals over time can also be sampled in an
MC procedure in which all the degrees of freedom are updated together efficiently \cite{Beard96,Prokofev98}. The relationships between the SSE
and continuous time formulation have been discussed in more detail in Ref.~\cite{Sandvik97}.

\vskip2mm
Like the SSE method, the perturbation expansion converges with a mean string length of order $\beta N$, regardless of the relative strengths of $H_0$ and $V$.
When $H_0$ dominates the energy the prefactor will be smaller in the perturbation expansion, which then is more economical. When
$H_0$ is not very dominant (which is common with quantum spin models, where the diagonal and off-diagonal energies are often similar in magnitude) the SSE
approach has an advantage owing to the fact that it is formulated in a completely discrete representation, while in the continuous-time formulation
floating-point numbers (the imaginary time points) have to be processed. Recently, the SSE scheme was also further developed for systems where $H_0$
dominates \cite{Albash17} (which often correspond to a nearly classical statistical system), essentially by integrating out all the diagonal terms from
the SSE operator strings. This approach may be as efficient as the continuous-time WL approaches (or even more efficient in some cases) when there
is a large imbalance in the strengths of the diagonal and off-diagonal terms. The original SSE approach should still be better (because of
the simplicity of the algorithm) when the terms are similar in strength.

\subsection{Stochastic series expansion for ground states}
\label{sec:projector}

\index{ground-state projector method}
In order to study the ground state of a system, one can take the limit $T \to 0$ within one of the methods discussed above. In practice, this means
$T \ll \Delta$, where $\Delta$ is the smallest finite-size excitation gap of the system. For systems with gaps closing as $1/L$ (where from now on $L$
will refer to the linear size of the lattice), the ground state can be routinely reached on systems with thousands, in many cases even tens of
thousands, of spins or bosons.

\vskip2mm
Ground states can also be studied using ``projector methods'', where, instead of tracing over a complete basis, the imaginary-time evolution operator
${\rm e}^{-\beta H}$ propagates some initial state $|\Psi(0)\rangle$ (often called ``trial state'', though this term can be misleading, since the final
result should be independent of the choice of initial state) that overlaps with the ground state; $|\Psi(\beta)\rangle = {\rm e}^{-\beta H} |\Psi(0)\rangle$.
By expanding in energy eigenstates, one can readily confirm that $|\Psi(\beta)\rangle$ approaches the ground state for sufficiently large $\beta$
(up to a normalization factor). 

\vskip2mm
In this case, the MC sampling is of terms contributing to the normalization $\langle \Psi(\beta)|\Psi(\beta)\rangle$. The numerator of an expectation value,
\begin{equation}
\langle A\rangle = \frac{\langle \Psi(\beta)|A|\Psi(\beta)\rangle}{\langle \Psi(\beta)|\Psi(\beta)\rangle},
\label{ags}
\end{equation}
is similarly expressed to obtain estimators to be averaged in the MC process. Here one can proceed as in the path integral approach, by
introducing a discrete slicing $\Delta_\tau=\beta/L$ of ${\rm e}^{-\beta H}$ or taking the limit $\Delta_\tau=0$ as in the continuous time formulation.
This type of method is known generically as the {\it path integral ground state} (PIGS) approach; \index{path integral ground state method}
for a review see Ref.~\cite{Yan17}. One can also
proceed with the ground state expectation value (\ref{ags}) as in the SSE method by series expansion and sampling strings of operators. In either case,
the main difference from the $T>0$ methods is the boundary condition in the time direction---at $T>0$ we have periodic boundaries, reflecting the trace
oparetion, while in the ground state approach the boundary condition is dictated by the starting state $|\Psi(0)\rangle$. This state is normally chosen such
that the time boundary condition is convenient for the updating process used in the MC sampling \cite{Sandvik10a}, and it can also some times be optimized
in some way, so that it already is a good approximation to the ground state. Note that the projection procedure is variational, so that $\langle H\rangle$
always approaches the true ground state energy monotonically from above.

\vskip2mm
Instead of projecting out the ground state with ${\rm e}^{-\beta H}$, a high power $(-H)^m$ of the Hamiltonian can also be used (where the negative sign is
just included for convenience as the ground state energy is normally negative). For sufficiently large $m$, $(-H)^m|\Psi(0)\rangle$ approaches the eigenstate
whose eigenvalue is the largest in magnitude, i.e., either the ground state of $H$ or of $-H$. Convergence to the ground state of $H$
can be ensured by adding a suitable constant to $H$.

\vskip2mm
We can now ask, what is the more efficient approach, using ${\rm e}^{-\beta H}$ or $(-H)^m$? Proceeding as we did above when discussing the distribution of
expansion orders in the SSE, we an see that the power $m$ required to reach the ground state is related to the $\beta$ value required with the exponential
operators as $m \approx \beta|E_0|$, where $E_0$ is the total ground state energy (which has been discussed in detail in Ref.~\cite{Sen15}). As in the SSE
method, $(-H)^m$ is expanded out into strings of the terms of the Hamiltonian, and these are sampled along with the starting state. There is no major
difference between the two approaches, as the summation over $n$ required with the series expansion is accomplished essentially for free when using
the fixed string-length SSE approach, Eq.~(\ref{zsse3}). The sampling of the operator strings does not differ significantly between the two approaches.

\vskip2mm
The ground state projector method is particularly useful when starting states $|\Psi(0)\rangle$ can be used that are tailored to the sampling
method used for the operator sequences. For Heisenberg quantum spin models, good variational starting states can be written in the valence-bond
basis \cite{Lou07,Lin12}, i.e., the overcomplete basis of singlet pairs, and these are ideally matched with loop-algorithm sampling schemes \cite{Sandvik10a}.
The valence-bond states have total spin $S_{\rm tot}=0$ and momentum $0$, which is the ground state sector for the models where the approach is suitable. Fixing
these quantum numbers from the outset can help to reduce the projection time $\beta$ (or power $m$)

\vskip2mm
We will not discuss further any specifics of ground-state projector methods, but just note again that the differences with respect to $T>0$
methods are very minor, and typically it is very easy to change a program from one case to the other.

\subsection{Quantum annealing with generalized stochastic series expansion}

\index{quantum annealing}
In projector QMC calculations, results obtained from a projected state $|\Psi(\beta)\rangle$ at first sight has no obvious use if the projection ``time''
$\beta$ is not sufficiently large for achieving convergence to the ground state. However, when considering ${\rm e}^{-\beta H}$ as a time evolution operator
in imaginary time, i.e., with time $t=-i\beta$ in $U(t)={\rm e}^{-itH}$, the projection corresponds to a quantum quench from the initial state $|\Psi(0)\rangle$
where at $t=0$ the Hamiltonian is suddenly changed from the one which has $|\Psi(0)\rangle$ as its ground state to whatever $H$ is considered in
the simulation. Though imaginary and real time evolutions are quite different, one can some time extract real-time dynamic information from such
imaginary-time quenches \cite{Weinberg17}.

\vskip2mm
Real-time quantum dynamical calculations are in general very difficult, with any existing method,
and it is interesting to ask what information may be gleaned from imaginary
time evolution, where, for sign free models, QMC calculations can be carried out for various out-of-equilibrium situations. The aforementioned quantum
quench in imaginary time is an extreme case of a more general setup where the Hamiltonian has some time dependence; $H=H(\tau)$ in imaginary time.
The time evolution operator, starting at $\tau=0$, is then
\begin{equation}
U(\tau) = T{\rm exp} \left ( -\int_0^\tau d\tau ' H(\tau ')    \right ),
\end{equation}
where $T$ indicates time-ordering, and one may consider expectation values
\begin{equation}
\langle A(\tau)\rangle = \frac{\langle \Psi(0)|U(\tau)AU(\tau)|\Psi(0)\rangle}{\langle \Psi(0)|U(\tau)U(\tau)|\Psi(0)\rangle}.
\label{atauevolved}
\end{equation}
Here one can again proceed with a time-slicing approach or apply a series expansion. Using the latter, the time evolution operator can be written as
\begin{equation}
U(\tau) = \sum_{n=0}^\infty (-1)^n \int_{\tau_{n-1}}^\beta  d\tau_n \cdots \int_{\tau_1}^{\beta} d\tau_2 \int_0^{\beta} d\tau_1
H(\tau_n) \cdots H(\tau_2) H(\tau_1),
\end{equation}
and one can proceed as in the several other cases discussed above and expand further into strings of terms of the Hamiltonian. Now, however,
the Hamiltonian depends on imaginary time and the string is always time ordered. One can sample the strings and the initial state $|\Psi(0)\rangle$ with
schemes very similar to the SSE method, and the the integrals can be taken into account by sampling sequences of ordered time points \cite{Degrandi11}.

\vskip2mm
There is also an alternative approach based on a product of Hamiltonians $H(\tau_m) \cdots H(\tau_2)H(\tau_1)$ for a fixed number of operators $m$, with
a fixed spacing between the time points and no integrals over time \cite{Liu13}. This results in a dynamics slightly different from the Schr\"odinger
dynamics in imaginary time, but for scaling purposes, e.g., when investigating dynamical quantum-criticality (i.e., the dependence on observables on the
rate at which the Hamiltonian is changed in time close to a critical point), the two approaches both give correct results. It should be noted that these
approaches really probe Hamiltonian quantum dynamics, and not the stochastic dynamics of the QMC sampling methods (as is the case with the other methods
often referred to as ``simulated quantum annealing'' \cite{Boixo14}) \cite{Degrandi13,Liu15}.

\section{Sampling algorithms and expectation values}
\label{sec:sampling}

We consider two classes of important $S=1/2$ quantum spin models to illustrate SSE sampling schemes. First, the Heisenberg antiferromagnet
defined by \index{Heisenberg antiferromagnet}
\begin{equation}
H = J \sum_{\langle ij\rangle} {\bf S}_{i} \cdot {\bf S}_{j} = J \sum_{\langle ij\rangle} [S^z_{i}S^z_{j} +
\hbox{$\frac{1}{2}$}(S^+_{i}S^-_{j} + S^-_{i}S^+_{j})],
\label{heisenberg}
\end{equation}
where $\langle ij\rangle$ refers to the nearest-neighbor sites pairs on an arbitrary bipartite (to avoid sign problems) lattice.
Second, we consider the transverse-field Ising model (which will hereafter be referred to as just the Ising model), which is often defined with
Pauli matrices, \index{transverse-field Ising model}
\begin{equation}
H = -\sum_{ij}J_{ij} \sigma^z_{i} \sigma^z_{j}  - h \sum_{i} (\sigma^+_{i}  + \sigma^-_{i}). 
\label{ising}
\end{equation}
Here the Hamiltonian is written with a generic Ising coupling $J_{ij}$ with no restriction on the range of the interactions, and we will
consider both short-range and long-range cases. The SSE method can also be used with long-range interactions in the Heisenberg case, though
with the limitation that there can be no frustration in order to maintain positive-definite sampling weights. With anisotropic Heisenberg
interactions, which we will consider later, the diagonal part can also be frustrated.
It is some times better to study the Ising model in a rotated basis, or, equivalently, to use the same $\sigma^z$ basis as above
but with the Hamiltonian written as
\begin{equation}
H = -\sum_{ij}J_{ij} \sigma^x_{i} \sigma^x_{j}  - h \sum_{i} \sigma^z_{i}, 
\label{ising2}
\end{equation}
where $\sigma^x_i = \sigma^+_i + \sigma^-_i$. Whichever version is better depends on details of the system (e.g., whether the interactions are frustrated,
if disorder is present, etc.) and what physical observables are to be calculated (noting that diagonal observables are often easier to access,
as we will see in sec.~\ref{sec:estimators}). We will here only discuss implementations of algorithms with the version in Eq.~(\ref{ising}).

\subsection{Configuration representations and diagonal updates}

As discussed in Sec.~\ref{sec:sse1}, a configuration in the SSE algorithm comprises a state $|\alpha_0\rangle$ and an index sequence $S_M = b_1,b_2,\ldots,b_M$,
the latter referring to a product of operators (and we will use the terms ``operators'' and ``indices'' interchangeably). Here we take the expression of the
partition function in Eq.~(\ref{zsse3}), where the index sequence is of a fixed length $M$ (whic can be determined in a self-consistent way by the program
during the equlibration part of a simulation). Removing the redundant complete sets of states we have
\begin{equation}
 Z_{\rm SSE} = \sum_{S_M} \frac{\beta^n(M-n)!}{M!} \sum_{\{ \alpha \}} \langle \alpha| H_{b_M} \cdots H_{b_2}H_{b_1}|\alpha\rangle ,
\label{zsse4}
\end{equation}
where in the products of $M$ operators there are $n$ terms of the Hamiltonian along with $M-n$ randomly distributed unit operators represented by the
index $b=0$. The number $n$ changes in updates that are called {\it diagonal updates}, because they involve replacing a unit operator $H_0$ by a diagonal
term of $H$ (which we simply refer to as an operator insertion), whence $n \to n+1$, or vice versa (an operator removal), in which case $n \to n-1$.
The generic way to carry out a sequence of diagonal updates is to go through the elements in $S_M$ one-by-one and to attempt a replacement of the
index whenever it does not correspond to an off-diagonal operator. 

\vskip2mm
Off-diagonal operators cannot be updated individually while maintaining the periodicity constraint $|\alpha(M)\rangle = |\alpha(0)\rangle$ for the propagated
states defined according to Eq.~(\ref{pstate}). How to carry out updates with the off-diagonal operator will be discussed in detail in the next section, but
here we note that the general strategy is to replace some number of diagonal operators by off-diagonal ones, or vice versa, without changing the lattice
location of the operator. With some models, the original Hamiltonian contains operators that makes this possible, e.g., in the case of the Heisenberg
interaction there are diagonal and off-diagonal operators on each bond, while in other cases certain constant diagonal operators have to be added
just for the purpose of enabling the necessary operator replacements. The operator replacements also imply changes in the states, and normally the SSE
algorithm is ergodic even if no other explicit state updates are included (though at high temperatures it can be useful to also carry out additional
updates only on the stored state $|\alpha\rangle$, keeping the operator string unchanged).

\vskip2mm
To carry out an update at position $p$ in the sequence requires the propagated state $|\alpha(p-1)\rangle = |\alpha(p)\rangle$ on which the diagonal operator
$H_{b_p}$ acts. The operator-index string is examined for each $p=1,\ldots,M$, and diagonal updates are attempted where possible. Whenever an off-diagonal operator
is encountered at a position $p$, the stored propagated state is advanced; $|\alpha(p)\rangle = H_{b_p}|\alpha(p-1)\rangle$. If an encountered index $b_p=0$,
one out of a number $N_d$ of diagonal operators can be chosen at random and inserted with a Metropolis acceptance probability, which depends on the matrix
element $\langle \alpha(p)|H_{b'_p}|\alpha(p)\rangle$, where $b'_p$ is the new generated diagonal index. If the current index $b_p \not=0$, a replacement with
$b'_p=0$ is attempted, and the acceptance probability then depends on the current matrix element $\langle \alpha(p)|H_{b_p}|\alpha(p)\rangle$. These updates
change the expansion power $n$ by $+1$ and $-1$, respectively, and this change also enters in the acceptance probability due to the factor $(M-n)!$ in
the configuration weigh in Eq.~(\ref{zsse4}). Importantly, to maintain detailed balance, the acceptance probabilities must also compensate for the inherent
imbalance stemming from the fact that there are $N_d$ ways of tentatively replacing an index $b_p=0$ by a non-zero index (some of which may correspond to vanishing
matrix elements, but that is irrelevant at this stage) when the number of Hamiltonian terms $n$ is increased by one, while for removing an operator there is
only one way of replacing $b_p \not=0$ by $0$.

\vskip2mm
With all the relevant factors and imbalance taken into account, the following are the correct generic acceptance probabilities for a single-operator
diagonal update with the SSE partition function (\ref{zsse4}):
\begin{subequations}
\begin{eqnarray}
  P(0 \to b_p) & = & \frac{\beta N_d \langle \alpha(p)|H_{b_p}|\alpha(p)\rangle}{M-n}, \label{pinsert} \\ 
  P(b_p \to 0)  & = & \frac{M-n+1}{\beta N_d \langle \alpha(p)|H_{b_p}|\alpha(p)\rangle}, \label{premove}
\end{eqnarray}
\end{subequations}
where $n$ is the number of Hamiltonian operators in $S_M$ before the update is carried out.
The matrix elements are trivial, and often some fraction of them vanish.

\vskip2mm
As an alternative to Metropolis-type updates as described above, one can carry out the diagonal updates according to a heat-bath scheme, where the relative
probabilities of all diagonal operators, including the fill-in unit operators, are considered and a choice is made among them according to their
relative probabilities (instead of choosing them with equal probability in the scheme above, irrespective of the values of the matrix elements). In many cases
it would take too long to compute these relative probabilities for each new state $|\alpha(p)\rangle$, but for sufficiently simple models it is possible to
avoid this step, at the cost of some rejected attempts, as will be discussed below in the context of systems with long-range interactions.

\begin{figure}[t!]
 \centering
 \includegraphics[width=0.9\textwidth]{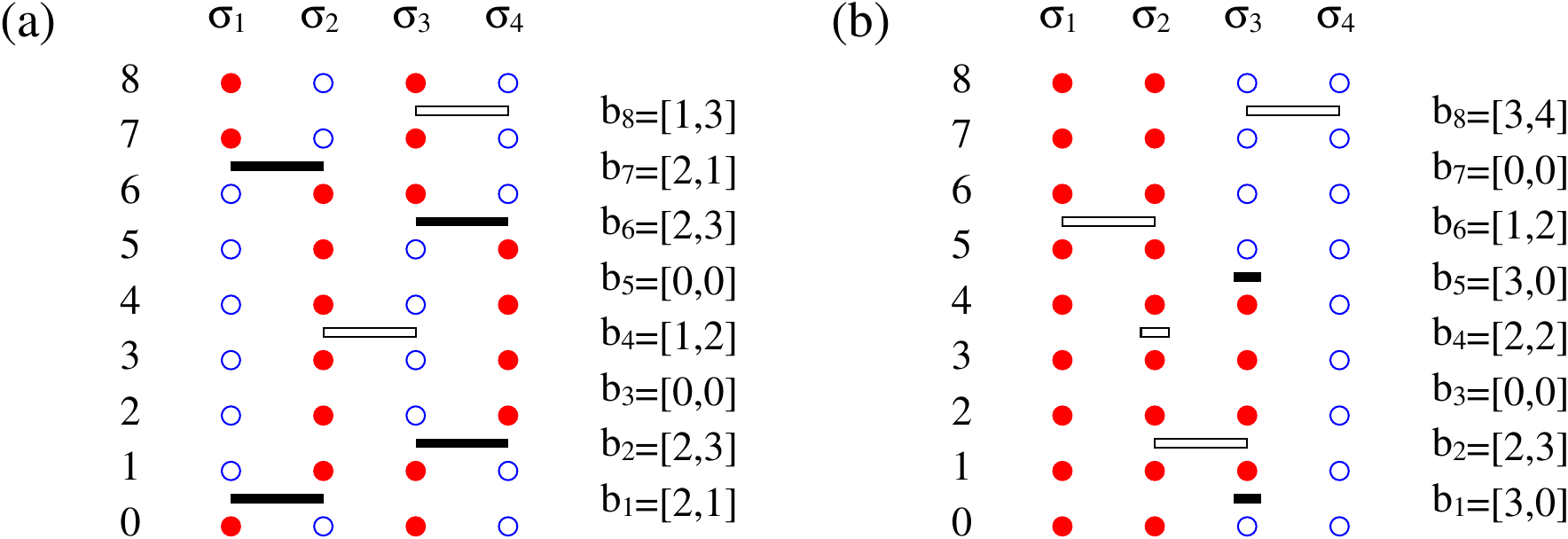}
 \caption{Graphical representations of SSE configurations for (a) the Heisenberg model and (b) the Ising model, in both cases for a system of four spins and
 with the SSE cutoff $M=8$. Up and down spins correspond to solid and open circles. All the propagated states $|\alpha(0)\rangle,\ldots,|\alpha(M)\rangle$,
 with  $|\alpha(M)\rangle=|\alpha(0)\rangle$, are shown along with the operators $H_{b_p}$. The number of Hamiltonian terms for both systems is $n=6$, and
 the two cases of empty slots between propagated states correspond to fill-in unit operators $H_{0,0}$ at these locations.  In (a) the solid and open bars
 represent, respectively, off-diagonal and diagonal parts of the Heisenberg exchange operators. In (b) the ferromagnetic Ising interactions are likewise
 represented by open bars, and the off-diagonal single-spin flip operators are represented by short solid bars. The short open bars correspond to the constant
 site-indexed operators.}
 \label{fig:config}
\end{figure}

\vskip2mm
Let us now be more specific and discuss how to represent the SSE configurations for the Heisenberg and Ising models. Illustrations are provided in
Fig.~\ref{fig:config}. We denote the spins in the stored state by $\sigma_i$, $i=1,2,\ldots,N$, and in a computer implementation they can be stored as integers,
$\sigma_i = \pm 1$. In both cases, to define the terms $H_b$ appearing in the operator products it is convenient to regard the subscript $b$ as formally
representing two indices (which can still for efficiency be packed back into a single integer in a program), but in slightly different ways for the
two models considered. In both cases we use the notation $H_{0,0}=1$ for the fill-in unit operators.

\vskip2mm
For the antiferromagnetic Heisenberg model (\ref{heisenberg}) we set $J=1$ and define the following diagonal and off-diagonal
operators:
\begin{equation}
H_{1,b} = \hbox{$\frac{1}{4}$} - S^z_{i}S^z_{j},~~~~~~ H_{2,b} = \hbox{$\frac{1}{2}$}(S^+_{i}S^-_{j} + S^-_{i}S^+_{j}).
\end{equation}
Then $H = -\sum_b [H_{1,b}-H_{2,b}]+N_b/4$, where $N_b$ is the total number of interacting spin pairs, e.g., $N_b=DN$ for $N$ spins on a $D$-dimensional
simple cubic lattice with periodic boundary conditions. Here a constant ${1}/{4}$ has been included in the diagonal operators, and they can therefore
act with a non-zero outcome only on two antiparallel spins. There is then a useful (as we will see) similarity with the off-diagonal terms, which also
can only act on antiparallel spins. The non-zero matrix elements of the Hamiltonian terms $H_{1,b}$ and $H_{2,b}$ are all $1/2$. The weight of an allowed SSE
configuration in Eq.~(\ref{zsse4}) is therefore $W(S_M) = (\beta/2)^n(M-n)!$, where the unimportant overall factor $1/M!$ has been omitted and there are
never any minus signs (for bipartite interactions) because the number of off-diagonal operators in the string has to be even.

\vskip2mm
Note that there is no explicit dependence of the weight on the state $|\alpha\rangle$ in Eq.~(\ref{zsse4}), but the state imposes constraints on the operator
string as only operations on antiparallel spins are allowed. An example of a very small Heisenberg SSE configuration is shown in Fig.~\ref{fig:config}(a). Note
again that the mean number of operators is $\propto \beta N$, and in large-scale simulations the number can be up to many millions.

\vskip2mm
In the diagonal update, if an encountered index pair at the current location $p$ is $b_p=[0,0]$, a bond index $b$ is generated at random among all the choices.
If the spins at the sites $i(b),j(b)$ connected by bond $b$ are antiparallel in the currently stored state $|\alpha(p)\rangle$, i.e., $\sigma_i\not=\sigma_j$, then
the operator $H_{1,b}$ is allowed and the index pair is set to $b_p=[1,b]$ with probability given by (\ref{pinsert}), where the matrix element equals $1/2$.
If the two spins are parallel nothing is changed and the process moves to the next position, $p \to p+1$. Each time an off-diagonal operator $[2,b]$ is
encountered, in which case no diagonal update can be carried out, the stored state is propagated; $\sigma_i \to -\sigma_i$ and $\sigma_j \to -\sigma_j$.

\vskip2mm
For the Ising model (\ref{ising}), where the Ising interactions $J_{ij}$ are of arbitrary range,
we define the following operators \cite{Sandvik03}:
\begin{equation}
H_{i,j} = |J_{ij}| -J_{ij}\sigma^z_{i}\sigma^z_{j} ~~~(i\not= j), ~~~~~ H_{i,i} = h, ~~~~~ H_{i,0}=h(\sigma^+_{i} + \sigma^-_{i}).
\label{isingops}
\end{equation}
Here the constant site-indexed operators $H_{i,i}$ serve as an example of how trivial diagonal terms can be added to the Hamiltonian for the purpose of carrying
out off-diagonal updates---as we will see in the next section, updates will be based on replacements $H_{ii} \leftrightarrow H_{i,0}$. In the diagonal updates,
the trivial constants will be inserted and removed along with the Ising operators $H_{i,j}$ ($i,j \not= 0$, $i \not=j$). In the Ising operators, the
presence of the constant $|J_{ij}|$ implies that only operation on a parallel pair of spins (for a ferromagnetic coupling $J_{ij}$) or an antiparallel
pair (for antiferromagnetic coupling) is allowed. This choice of the added constant is again motivated by its convenience for constructing the off-diagonal
(cluster) update, as we will see further below. The example of an SSE Ising configuration in Fig.~\ref{fig:config}(b) only includes nearest-neighbor
ferromagnetic interactions.

\vskip2mm
For a $D$-dimensional simple cubic lattice and only nearest-neighbor interactions $J$ included, there are now $N_d=(D+1)N$ diagonal operators, of which
$DN$ are Ising terms and $N$ are the constant operators $H_{i,i}$. When these are all generated with equal probability in attempts to insert operators,
with the Metropolis acceptance probability given by Eq.~(\ref{pinsert}), there is an inherent inefficiency if $h$ and $J$ are very different. For example, if
$J=1$ and $h \ll 1$, most of the attempted $h$-operator insertions will be rejected.

\vskip2mm
The rejection problem becomes worse with long-range interactions, e.g., \index{long-range interactions}
if $J_{ij}$ decays with the separation $r_{ij}$ as a power law, $|J_{ij}| \propto r_{ij}^{-\alpha}$. Then there are $N(N-1)/2+N=N(N+1)/2$ diagonal operators
to generate at random, and those with very small $J_{ij}$ will be rejected almost always. This problem can be overcome easily by generating the diagonal
Hamiltonian terms along with the fill-in operators $H_{00}$ using a heat-bath method \cite{Sandvik03}. Instead of treating operator insertions and
removals as different types of updates, these are now combined and carried out at all positions $p$ at which the current operator $H_{b_p}$ is not
off-diagonal. The method needs a precomputed table of integrated relative probabilities $P_{ij}$ of all the different diagonal operators, where it
is tentatively assumed that all operators are allowed. The probabilities are calculated from Eq.~(\ref{zsse4}) and the definitions in (\ref{isingops}),
and, for efficiency, mapped into a single-index ordered table $P_k$, $k=1,\ldots,N(N+1)/2$. In each diagonal update, a random number $ r \in [0,1)$ is
generated and the corresponding operator is identified in the table, using the bisection method to search in the ordered table for the corresponding
probability window $P_{k-1} \le r < P_k$, thus finding the correct operator indices $i(k),j(k)$ in $\propto \ln(N)$ operations. A new Ising operator has
to be rejected if the spins $\sigma_i,\sigma_j$ are incompatible with the sign of the interaction $J_{ij}$.

\vskip2mm
A very useful aspect of this approach is that it
renders an algorithm with processing time scaling as $N \ln(N)$ for long-range interactions, instead of the naively expected $N^2$ scaling; details are
described in Ref.~\cite{Sandvik03}. This efficient method for the diagonal updates can also be used for the Heisenberg model with long-range interactions,
and for many other cases as well. Even with short-range interactions, the heat-bath approach may be slightly more efficient than the Metropolis
update, though the difference in efficiency is likely minimal in most cases.

\vskip2mm
The probability of finding a $0$-element (fill-in operator) and attempting an operator insertion in the string clearly depends on how the cutoff
$M$ of the series expansion is chosen. The cutoff naturally should be high enough for the contributions from terms with $n>M$ to be completely negligible,
so that the truncation of the series expansion is no approximation in practice. In an SSE simulation this can be ensured by always requiring $M$ to be
significantly larger than the largest $n$ that has been reached so far during the equilibration part of the simulation (with any adjustments done after
each completed sweep of diagonal update during equilibration). In practice, $M$ exceeding the maximum $n$ by 30-50\% is a suitable chose; clearly
sufficient for causing no systematical error and also enough to allow a large number of operator insertion attempts. Normally $M$ (and $\langle n\rangle$)
converges very quickly at the initial stages of a simulation.

\subsection{Loop and cluster updates}

\vskip2mm
In classical MC simulations of spin models, cluster updates \cite{Swendsen87,Wolff89} have played a major role in reaching system sizes sufficiently large for
reliable finite-size scaling studies. These methods also have generalizations for some quantum spin models
\cite{Evertz93,Ying93,Kawashima94,Beard96,Sandvik99,Evertz03,Sandvik03},
including the Heisenberg and Ising systems discussed here. Within the SSE approach, the loop
and cluster updates are carried out in the operator string, which at this stage is regarded as a network of connected {\it vertices} comprising the
operators and their associated ``incoming'' and ``outgoing'' states (i.e., the information needed to compute the weight of the operator string).
The general strategy is to update a set of vertices (which are connected to each other to form a loop or a cluster) but maintain their lattice
locations (which dictate their connectivity). The connectivity is changed only as a consequence of the diagonal updates. The off-diagonal updates also
can change the stored state $|\alpha_0\rangle$, since the loops or clusters can span across the periodic time boundary
represented by the stored state.

\subsubsection{Linked vertex list}

The loop or cluster updates are carried out in the linked vertex list, which after a full sweep of updates is mapped back into the simple operator-index
string and state $|\alpha\rangle$ (or, in some cases, it is better to do this mapping-back continually during each loop or cluster flipping procedure).
A vertex comprises the spin states before and after an operator has acted, and these states are associated with the {\it legs} of the vertex,
e.g., for a two-body interactions there are four legs for each vertex; two before and two after the operator has acted on the two spins. Thus, the vertex
is associated with a matrix element of the form $\langle \sigma_i(p),\sigma_j(p)|H_{b_p}|\sigma_i(p-1),\sigma_j(p-1)\rangle$, where $i$ and $j$ are the
sites involved in the interaction term $H_{b_p}$. The way the legs of different vertices are linked to each other corresponds directly to the segments of
unchanged spins between operators in the ``time'' direction in Fig.~\ref{fig:config}. These lines of spins are redundant when representing the changes
that can be made in a configuration by changing the type of some of the operators (e.g., diagonal to off-diagonal or vice versa) and making associated changes
in the spin state (represented by the changes at the vertex legs). In a computer program, these lines of constant spins are represented by bidirectional
links connecting the relevant vertices, enabling direct jumps between any of the connected vertex legs without regard for the intermediate
irrelevant spins. The linked list can be constructed according to a simple and efficient scheme discussed in detail in Ref.~\cite{Sandvik10b}.

\subsubsection{Loop and cluster construction}

\begin{figure}[t!]
\centering
\includegraphics[width=0.72\textwidth]{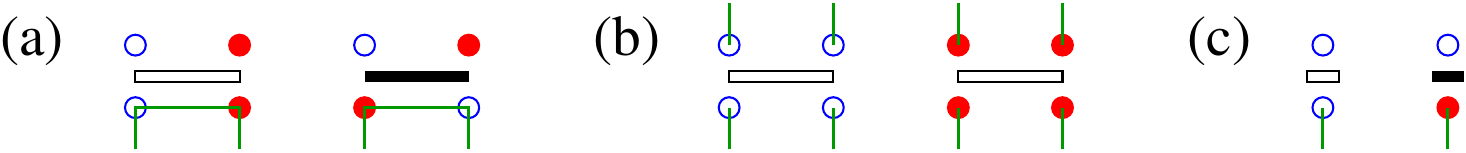}
\caption{Examples of the elementary vertex processes by which loop and cluster updates are carried out for Heisenberg and Ising models.
The green line segments represent parts of a loop or cluster. The spins and operators on a given vertex correspond to the state before the loop
[in (a)] or cluster  [in (c) and (d)] has been flipped, and the adjacent vertex shows the state after the loop or cluster has been flipped.}
\label{fig:process}
\end{figure}

\index{cluster update}\index{loop update}
Some relevant vertices are illustrated in Fig.~\ref{fig:process}, along with lines indicating how the legs of vertices can be affected by a loop or cluster
flip. The general idea is to change the spin state at one vertex leg and then move either to another leg of the same vertex or follow a link to another
vertex, until the process closes. The allowed processes depend on the Hamiltonian. In the simplest cases, the {\it deterministic loops}
for Heisenberg models and Ising cluster updates, there is a unique way to move at each step. The entire system can then be divided into a unique set of
loops or clusters, each of which can be flipped (i.e., changing the spin states at the affected vertex legs) without affecting the configuration weight.
Each loop or cluster can then be flipped independently with probability $1/2$, as in the classical Swendsen-Wang algorithm.

\vskip2mm
The condition that the loop or cluster flip must not change the configuration weight is the main
restriction of this type of update, and is very similar to the limitation of classical cluster updates, e.g., the Swendsen-Wang cluster update for
the Ising model \cite{Swendsen87} does not work in the presence of an external magnetic field. In the next section we will discuss more complicated
{\it directed loop updates}, which partially overcome this limitation.

\begin{figure}[t!]
 \centering
 \includegraphics[width=0.95\textwidth]{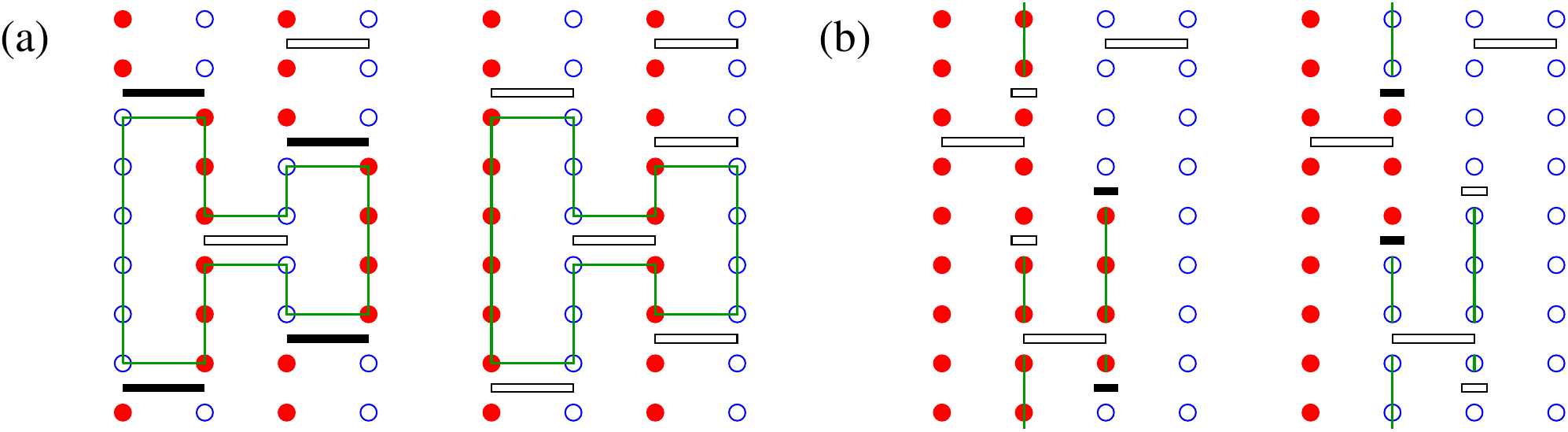}
 \caption{Examples of entire loop (a) and cluster (b) updates based on the elementary processes shown in Fig.~\ref{fig:process}. The
 pairs of configurations correspond to the states before and after the loop or cluster (indicated by green lines) has been flipped. Note
 that in (b) the cluster spans across the periodic time boundary. In  cases where the loop and cluster algorithms are efficient, the vertex weight
 (matrix element) is the same before and after the flip.}
 \label{fig:updates}
\end{figure}

\vskip2mm
We will not discuss the details of the SSE loop \cite{Sandvik99,Sandvik10b} and cluster updates \cite{Sandvik03} here, but refer to the literature.
We just note that the loop update, also called {\it operator-loop} update within the SSE framework to emphasize that everything is formulated with
a network of connected vertices (operators), corresponds to moving along a one-dimensional, non-branching path in the network of vertices. At
some point, this process necessarily will lead back to the vertex leg at which the process started, at which point the loop is completed and a new
one can be started. An example of a loop is shown in Fig.~\ref{fig:updates}(a). Normally these loops are constructed one-by-one, with the random decision
of whether or not to flip the loop made before each new loop is started, and in each case the traversed vertex legs are flagged as visited, so that
each new loop can be started from a not previously visited leg until all legs have been visited.

\vskip2mm
The cluster update for the Ising model differs from the Heisenberg loop update in the important sense of being branching. Since each two-body vertex must
have all four legs in the same spin state, because of the choice of the added constant in the Ising operators $H_{ij}$ in Eq.~(\ref{isingops}),
once one of those spins has been flipped the other three must be flipped as well to ``heal'' the vertex. The cluster therefore can branch out each time a
two-body vertex is encountered. However, no branches extend out from vertex legs that have been previously visited, and eventually the cluster therefore
does not grow further, and all defective vertices generated in the process have been healed. Branches that point to a single-site operator also terminate,
because by flipping the type of single-site vertex from $H_{i,0}$ to $H_{i,i}$, or vice versa, the propagating defect is healed and not further changes
need to be made at that branch. Fig.~\ref{fig:updates}(b) shows one example, which in this small configuration involves a single two-body operator and
four one-body operators. In general, the branching-out can propagate to a very large number of vertices.

\vskip2mm
In a program implementation, the successive branching-out is handled by adding new branches to a stack, from which branches to follow are picked
one-by-one until the stack is empty (the cluster is completed).
By starting each new cluster from a vertex leg not previously visited, the entire system will eventually
be subdivided into clusters (each of which was flipped by probability $1/2$), as in the Swendsen-Wang algorithm.

\subsubsection{Directed-loop updates}

\index{directed loop update}
The loop update described above can be generalized to Heisenberg systems with anisotropic interactions
and uniform magnetic fields, i.e., pair interactions of the form
\begin{equation}
H_{ij} = J_{ij} [\Delta_{ij} S^z_iS^z_j + \hbox{$\frac{1}{2}$}(S^+_iS^-_j + S^-_iS^+_j) + h_iS^z_i + h_jS^z_j],
\end{equation}
to which a suitable negative constant may have to be added in order to avoid a sign problem originating from the diagonal terms. In the directed-loop
algorithm \cite{Syljuasen02}, the one-dimensional path through the vertex list is no longer unique, in contrast to the deterministic loop update,
but a choice on how to continue the path has to be made at each vertex. A set of equations, called the {\it directed-loop equations} relate the
possible vertex weights to the probabilities of the different paths through a vertex, and by solving these equations a process maintaining detailed
balance is ensured.

\vskip2mm
The paths through the system taken in the directed-loop algorithm are similar to those in the continuous-time worm algorithm \cite{Prokofev98}, but the
probabilities are different. The directed loops often lead to a much more efficient evolution of the configuration. The directed loops can also be directly
formulated within the continuous-time framework \cite{Syljuasen02}, and implementations of continuous- and discrete-time WL methods now often rely on
the directed loop ideas \cite{Gubernatis16}.

\subsubsection{Sweeping cluster update}

\index{sweeping cluster update}
A very interesting recent development of efficient updates within the SSE approach is the {\it sweeping cluster update} developed for highly
constrained models such as the quantum dimer model \cite{Yan19}. It is somewhat similar to an earlier {\it multibranch cluster update} developed
in order to enhance the performance of a loop algorithm for a bosonic model with a constrained ring-exchange interaction \cite{Melko05}, but is more
directly tailored to strict geometric restrictions. Simulation results for the square-lattice quantum dimer model indicate that this is a promising
approach, overcoming at least some of the rather severe limitations of previous methods for this important class of models.

\subsubsection{Extended-cell approach}

\index{extended-cell expansion}
In the original SSE formulation discussed in Sec.~\ref{sec:sse1}, the terms $H_b$ are defined as they appear in the Hamiltonian, e.g., they are the single-site
operators such as $S^+_i+S^-_i$ or two-body terms such as $S^+_iS^-_j+S^-_iS^+_j$. A greater degree of freedom can be given to loop and cluster updates by 
enlarging the cell on which the operators and vertices are defined  \cite{Louis04}. For instance, one can define an off-diagonal operator
on three sites as $(S^+_iS^-_j+S^-_iS^+_j)I_k$, where $I_k$ is a unit operator with a site index. This allows the off-diagonal part to move during an update, e.g.,
a vertex with the mentioned operator can be updated (within some scheme involving more than one operator) to $(S^+_iS^-_k+S^-_iS^+_k)I_j$, This trick has
proved to be very helpful for speeding up SSE simulations of systems where the diagonal interactions are highly
frustrated \cite{Heidarian05,Isakov06,Melko07,Biswas16}.

\subsubsection{Loop and cluster updates at $T=0$}

As we saw in Sec.~\ref{sec:projector}, the $T>0$ SSE algorithm can be very easily modified for ground-state projection, with the only essential difference
being the change from periodic time boundary conditions to boundary conditions dictated by the trial state. How this change affects the loop and
cluster algorithms depends on the type of trial state used.

\vskip2mm
In the case of spin-isotropic Heisenberg models, a particularly good choice of trial state is one written in the valence-bond basis, i.e., the overcomplete
basis of singlet pairs, with the two members of each singlet occupying different sublattices on a bipartite lattice \cite{Sandvik10a}. The valence bonds
then serve as segments of loops touching the trial state, and there is no weight change, as before, when flipping a loop. The valence-bond trial state can
also in many cases be variationally optimized, for improved convergence \cite{Lou07,Lin12}. Typically the valence-bond trial state is a superposition of
valence-bond coverings, and the simulation involves a simple update for reconfiguring the bonds.

\vskip2mm
A convenient type of trial state to use in combination with the Ising cluster update is the ferromagnetic product state in the $x$ spin direction,
i.e.,  $|\Psi(0)\rangle =\prod (\uparrow_i +\downarrow_i)$, where $\uparrow$ and $\downarrow$ correspond to $\sigma^z_i=\pm 1$. This state corresponds to
a completely open time boundary condition, where cluster branches simply terminate at the trial state, and boundary clusters can be flipped as any
other cluster without any change to the configuration weight. At the opposite extreme, it may some times be useful to use a fully polarized ferromagnetic
state as the trial state \cite{Liu13}, in which case any cluster touching the boundary cannot be flipped at all.

\section{Estimators for expectation values}
\label{sec:estimators}
        
Expectation values suitable for estimation with MC simulation are normally written in the form
\begin{equation}
\langle A \rangle = \frac{\sum_C A_CP_C}{\sum P_C},
\label{aexpmc}
\end{equation}
where $\{ C\}$ is some configuration space, $P_C$ is the probability (or relative probability) of configuration $C$, and $A_C$ is the
estimator of the quantity $A$. In the QMC algorithms discussed in the preceding sections the sum of weights or probabilities in the denominator
is the partition function at $T>0$ and the wave-function norm in $T=0$ projector methods. When importance-sampling the configurations, so that
their probability of being generated equals $P_C$, the simulation result for $\langle A\rangle$ is the mean value  $\langle A_C\rangle$, with a
statistical error computed using some data binning technique. Here we discuss the form of the estimator $A_C$ for some important classes of
physical observables.

\vskip2mm
We should first note that, in classical MC simulations, $A_C$ is normally just a trivial function of the sampled degrees of freedom, and that is
also the case with SSE and WL methods when the operator $A$ is diagonal in the basis used. However, for off-diagonal operators the situation is more
complicated, as then the configurations $C$ contributing to $\langle A\rangle$ are not necessarily the same as those that contribute to the sampled
normalization in Eq.~(\ref{aexpmc}). We will here first discuss diagonal operators and then briefly touch on the topic of off-diagonal correlation
functions.

\subsection{Diagonal correlations and susceptibilities}

\index{susceptibility}
Equal-time expectation values, e.g., two-point or multi-point correlation functions, of diagonal operator in the basis used are normally
trivial, and for improved statistics they can be averaged over the SSE propagation index $p$;
\begin{equation}
\langle A\rangle = \left \langle \frac{1}{n} \sum_{p=0}^{n-1} A(p) \right \rangle,
\label{averageoverp}
\end{equation}
where $A(p)=\langle \alpha(p)|A|\alpha(p)\rangle$ is the value of $A$ computed in propagated state $p$ in a given SSE configuration. Here when defining the
propagated state $|\alpha(p)\rangle$ as in Eq.(\ref{pstate}) we have implicitly disregarded the fill-in unit operators $H_0$ in the formulation of the
method with a fixed string length $M$. It is also correct to include the unit operators and average over $M$ propagated states. In many cases it is too
time consuming to average over all states, and since the propagated states are highly correlated there is also no loss of statistics in practice to
average over a small fraction of the states, e.g., only including every $N$th or so state in Eq.~(\ref{averageoverp}). Similarly, spatial averaging
may be carried out fully or partially to improve the statistics.

\vskip2mm
We are often interested in correlation functions in Fourier space of a periodic system, e.g., with the transformed diagonal spin operator
\begin{equation}
S^z_q = \frac{1}{\sqrt{N}} \sum_{\bf r} {\rm e}^{-{\bf q}\cdot {\bf r}} S^z_{\bf r}.
\label{szq}
\end{equation}
The real part of the correlation function (often called the structure factor) is $S({\bf q}) = \langle S^z_{-q}S^z_q\rangle$ and the imaginary part
vanishes by symmetry. If all or many values of the wave-vector ${\bf q}$ are needed, the most efficient way is often to evaluate (\ref{szq}) using the
{\it Fast Fourier Transformation} method (see, for example, Ref.~\cite{Numrec}). However, if only a small number of values are needed it can be
better to use the basic summation formula just for those cases.

\vskip2mm
Time dependent (in imaginary time) correlations can also be calculated, and by numerical analytic continuation they can provide real-frequency
dynamic spectral functions of experimental interest, e.g., the dynamic spin structure factor measured in inelastic neutron scattering (see Ref.~\cite{Shao17}
for a recent discussion of of analytic continuation and spectral functions). The way imaginary time is treated within the SSE method was already
considered in Eq.~(\ref{atau}); a fixed value $\tau$ of imaginary time corresponds to a summation over separations $p$ between propagated states.
Because the weighs of the different separations follow a very narrow distribution, only a small fraction of them has to be summed over in practice,
and one can use a precomputed table of safe lower and upper bounds on the separations for the different $\tau$ values considered.

\vskip2mm
Alternatively, time dependent quantities can be more easily calculated if the SSE is formulated in combination with time slicing, where
first the exponential operator is written as $({\rm e}^{-\Delta H})^m$, with $\Delta=\beta/m$, and the factors are series expanded individually
with equal cutoffs $M_\Delta$. This is no approximation, since the slicing does not involve any separation of non-commuting operators. The
operator string consists of $m$ segments, $i=1,\ldots,m$, containing $n_i$ Hamiltonian terms. Therefore, in the diagonal updates, $\beta$
and $n$ in Eqs.~(\ref{pinsert}) and (\ref{premove}) are replaced by $\Delta$ and $n_i$, with $i$ corresponding to the segment in which an update
is carried out. The off-diagonal updates are not affected at all by the slicing, since they do not change the numbers $n_i$. Time-dependent
correlations of diagonal operators can now be evaluated easily by just considering the propagated states at the boundaries between slices,
which correspond to sharp time displacements in multiples of $\Delta$.

\vskip2mm
An interesting aspect of the SSE method is that rather simple estimators for static susceptibilities can be constructed. Such susceptibilities are
in general given by Kubo integrals of the form
\begin{equation}
\chi_{AB} = \int_0^\beta d\tau \langle A(\tau) B(0) \rangle ,
\label{chikubo}
\end{equation}
where $h_A A$ can be regarded as a perturbation added to the Hamiltonian (the number $h_A$ being a corresponding field strength) and $\langle B\rangle$
is the linear response of the operator $B$ to this perturbation; $\langle B\rangle = h_A \chi_{AB}$. If $A$ and $B$ commute with $H$, then
$\chi_{AB} = \beta \langle AB\rangle$, an important
example of which is the uniform magnetic susceptibility of a system in which the magnetization $M$ is conserved; then $A=B=M$ and $\chi_u=\beta\langle M^2\rangle$
as in a classical system. In the more common case
where $A$ and $B$ do not commute, e.g., if they are individual spins at different sites; $A=S^z_i$, $B=S^z_j$, the integral over the time dependent correlation
function has to be evaluated. Within SSE, the integral can be computed for each individual SSE configuration, with the result \cite{Sandvik91,Sandvik92}
\begin{equation}
  \chi_{AB} = \left \langle \frac{\beta}{n(n+1)} \left [ \left (\sum_{p=0}^{n-1} A(p) \right )\left (\sum_{p=0}^{n-1} B(p) \right )+ \sum_{p=0}^{n-1} A(p)B(p) \right ] \right \rangle,
\end{equation}
where $A(p)$ and $B(p)$ are the eigenvalues of the respective diagonal operators computed in propagated state $p$. Often these susceptibilities are also
needed in Fourier space, but it can still be better to do the computations in real space (depending again on how many points in momentum space are
needed, etc.) and only take the Fourier transform as the last step on the final averaged real-space susceptibilities.

\subsection{Off-diagonal observables}

We have already encountered two estimators for off-diagonal observables in the SSE method; the total energy and heat capacity
\begin{equation}
E=\langle H\rangle = -\frac{\langle n\rangle}{\beta},~~~~~~~ C=\frac{dE(T)}{dT} = \langle n^2\rangle - \langle n\rangle^2 - \langle n\rangle,
\label{eexpval}
\end{equation}
which depend only on the number $n$ of Hamiltonian terms in the sampled operator sequences. These expressions are exactly the same as
those in Handscomb's method. Note that the internal energy $E$ contains any constants that have been added to $H$ for the sampling scheme.

\vskip2mm
For any operator $H_b$ in the Hamiltonian, diagonal or off-diagonal, its expectation value is simply given by
\begin{equation}
\langle H_b\rangle = -\frac{\langle n_b\rangle}{\beta},
\end{equation}
where $n_b$ is the number of instances of the index $b$ in the sampled sequences $S_M$. Thus, the energy estimator in Eq.~(\ref{eexpval})
corresponds to summing over all $b$. The simplicity of off-diagonal observables that can be related to the terms in the Hamiltonian also carries
over to correlation functions \cite{Sandvik97};
\begin{equation}
\langle H_aH_b\rangle = \frac{(n-1) \langle n_{ab} \rangle}{\beta^2},
\end{equation}
where $n_{ab}$ is the number of times the indices $a$ and $b$ appear next to each other in $S_n$, where $S_n$ is the sequence obtained
from $S_M$ when all zero indices are disregarded. This result can also be generalized to time dependent correlations.

\vskip2mm
Kubo integrals of the form (\ref{chikubo}) also have their direct SSE estimators. For $A=H_A$ and $B=H_B$ (terms in the Hamiltonian corresponding
to the operators $H_b$ in the formalism above), after integrating over time for each SSE configuration we have
\cite{Sandvik92}
\begin{equation}
\chi_{AB} =\beta^{-1}(\langle n_An_B \rangle - \delta_{AB} \langle n_A\rangle ),
\label{chiaboff}
\end{equation}
which is very easy to evaluate. Two important off-diagonal quantities of this type can be mentioned (see the cited references for details):

\vskip2mm \index{spin stiffness}
The spin stiffness is the second-order energy (or free energy at $T>0$) response to a boundary twist imposed on the spins, or, alternatively, to a continuous
twist field analogous to a vector potential. The spin stiffness is also analogous to the superfluid stiffness of a boson system. The stiffness (spin or
superfluid) has an interesting estimator related to fluctuations of winding numbers \cite{Pollock87}, essentially the currents circulating around a
periodic system. The SSE estimator for the stiffness is similar to Eq.~(\ref{chiaboff}), involving the counting of off-diagonal operators transporting spin
(or charge) to ``left'' or ``right'' \cite{Sandvik97b}.

\vskip2mm
\index{fidelity susceptibility}
Recently, a simple algorithm for computing the fidelity susceptibility was proposed \cite{Wang15} (likely more efficient than a different
method proposed earlier \cite{Schwandt09}). This observable, which quantifies at what rate a state changes when some parameter is varied, plays an important
role in quantum information theory and is also useful in studies of quantum phase transitions. The estimator for the fidelity susceptibility is similar
to a correlator of the form (\ref{chiaboff}), with $n_A$ and $n_B$ corresponding to operators related to the infinitesimally varied coupling constant.
These operators are counted in different halves of the time-periodic SSE configurations.

\vskip2mm
\index{entanglement entropy}
Two other SSE-accessible information-theory inspired quantities can also be mentioned; the {\it entanglement entropy} \cite{Hastings10} and the
{\it mutual information} \cite{Melko10}, \index{mutual information}
both in their Rennyi versions. They are evaluated using a so-called swap operator, or a modified
space-time lattice corresponding to a ratio of two different partition functions.

\vskip2mm
To evaluate general off-diagonal correlation functions, which cannot be expressed simply with the terms of the Hamiltonian, one has to go outside
the space of the terms contributing to the partition function (or wave function norm in $T=0$ approach). An efficient scheme was first devised within
the worm algorithm \cite{Prokofev98}, and a simple generalization to the SSE framework is also known \cite{Dorneich01}. We will not discuss details
here, but just note that in the context of loop, directed-loop, or worm algorithms, the loop or worm building process involves two point defects 
that can be associated with the raising or lowering operators, $S^+_i$ and $S^-_j$ (or creation and destruction operators in particle models).
Space-time correlation functions involving these operators, e.g., $\langle S^+_i(\tau)S^-_j(0)\rangle$ are therefore directly related to the
probability distribution of the separation between the defects.

\section{Recent applications}
\label{sec:examples}

MC simulations have played, and continue to play, an important role in studies of phase transitions in classical statistical physics. In a similar
way, QMC simulations of quantum lattice models are now helping to push the boundaries of knowledge in the field of quantum many-body physics, uncovering
various quantum states and elucidating the nature of various quantum phase transitions (i.e., transitions between different types of ground states
and associated scaling behaviors at $T>0$) using sign free ``designer Hamiltonians'' \cite{Kaul13}. \index{designer Hamiltonian}
Some selected applications of SSE methods to different
classes of quantum-critical models will be very briefly reviewed in Sec.~\ref{sec:critical}, as a guide to recent works and mainly reflecting the author's
own interests. Works aimed at extracting dynamic spectral functions from imaginary-time correlations, using numerical analytic continuation methods,
will be discussed in Sec.~\ref{sec:dynamics}. Finally, SSE works on disorder (randomness) effects are reviewed in Sec.~\ref{sec:disorder}.
The emphasis is on SSE applications, and references to works using other methods are therefore very incomplete.

\subsection{Quantum phases and criticality in spin systems}
\label{sec:critical}

\index{dimerized Heisenberg model}\index{quantum criticality}\index{Heisenberg antiferromagnet}
One of the first successes of WL-type QMC simulations in quantum magnetism was the convincing demonstration of long-range order at $T=0$ in the 2D $S=1/2$
Heisenberg antiferromagnet \cite{Reger88} (the currently most precise results were obtained with the SSE \cite{Sandvik97b} and valence-bond projector
\cite{Sandvik10a} methods). Following this important result (of which there is still no rigorous analytical proof), the focus shifted to ways in which
the long-range order can be destroyed by perturbing the Heisenberg model in various ways. Many studies were devoted to statically dimerized 2D Heisenberg
models, e.g., the SSE studies in Refs.~\cite{Sandvik94,Wang06,Wenzel08,Ma18}, where there is a pattern of nearest-neighbor couplings of two strengths,
$J_1$ (inter-dimer) and $J_2$ (intra-dimer), such that each spin belongs uniquely to a dimer. As a function of the ratio $g=J_2/J_1$, there is then a loss
of antiferromagnetic order at $T=0$ when $g$ reaches a critical value $g_c$. The physical mechanism of this transition is that the density of singlets on
the dimer increases with $g$, and eventually the ground state becomes close to a product state of dimer singlets. The transition is continuous and belongs
to the 3D O(3) universality class, with the third dimension corresponding to imaginary time. In some cases, confirming this universality was challenging
\cite{Wenzel08}, because of, as it turns out \cite{Ma18}, effects of competing scaling corrections. The $T=0$ quantum-critical point is associated with a
so-called {\it quantum-critical fan} extending out in the $(T,g)$ plane from $g_c$ and which is associated with various scaling laws of physical
quantities \cite{Chubukov94}. SSE and other QMC studies have, among other things, established the rangle of validity of these asymptotic scaling behaviors,
and also tested the applicability of various approximate analytical calculations \cite{Sen15}, e.g., the $1/N$ expansion, where $N$ is the number of
spin components.

\vskip2mm
\index{O(3) transition}
The O(3) transition driven by dimerization can be realized experimentally in the 3D spin-dimer system TlCuCl$_3$ under pressure \cite{Merchant14}
and this has motivated SSE simulations of this phase transition also in 3D generalizations of the 2D Heisenberg systems discussed above. In a 3D Heisenberg
system antiferromagnetic long-range order can survive also at $T>0$ (which is excluded by the Mermin-Wagner theorem in 2D). An empirical universal scaling
form of the critical temperature was found in Ref.~\cite{Jin12} and further studied in Ref.~\cite{Tan18}. Multiplicative logarithmic corrections at
the $T=0$ and $T>0$ phase transitions have also been  studied in detail \cite{Qin15}
  
\vskip2mm

In the statically dimerized 2D and 3D systems, the paramagnetic phase is a unique quantum state with no spontaneous symmetry breaking---the singlets simply
form, with some fluctuations, at the dimers imposed by the Hamiltonian itself. A more interesting case is where also the paramagnetic state breaks
additional symmetries spontaneously. It was discovered by SSE simulations that a certain planar [XY, or U(1) symmetric] $S=1/2$ spin model could
go through a transition from XY magnetized to spontaneously dimerized in what appeared to be a continuous quantum phase transition \cite{Sandvik02a}.
Shortly thereafter,
a theory was proposed for a new type of quantum phase transition, beyond the Landau-Ginzburg-Wilson (LGW) paradigm, between 2D quantum antiferromagnetic
and spontaneously dimerized states (also called valence-bond-solids, VBSs). \index{valence-bond solid}
In this theory of {\it deconfined quantum critical points} \cite{Senthil04}, \index{deconfined quantum-critical point}
the two different order parameters arise out of the same objects---spinons and gauge fields---instead of being described by separate fields corresponding to
the two order parameters. This theory stimulated many further works on various 2D quantum antiferromagnets with VBS transitions, and these
studies have uncovered a rich variety of phenomena beyond the original DQCP proposal.

\vskip2mm
Traditionally, VBS states were discussed in the context of frustrated Heisenberg models, such as the model with first $J_1$ and second $J_2$ neighbor
interactions on the square lattice. These models have sign problems and are not accessible to QMC simulations. Another class of models, was proposed to
study the DQCP phenomenon with QMC without sign problems---the $J$-$Q$ model, \index{J-Q model} in which the Heisenberg exchange is supplemented by a
multi-spin interactions built out of $S=1/2$ singlet operators $P_{ij} = 1/4 - {\bf S}_i\cdot {\bf S}_j$ \cite{Sandvik07}. This interaction by itself,
$-JP_{ij}$, is equivalent to the antiferromagnetic Heisenberg echange. Products of two or more of the singlet projectors make up the competing $Q$ interaction,
e.g., terms of the form $-QP_{ij}P_{kl}$ with the involved sites $i,j,k,l$ suitably arranged on plaquettes (without breaking the symmetry of the square lattice).

\vskip2mm
When the ratio $Q/J$ is large, the correlated singlets favored by the multi-spin interactions cause
many of these models undergo quantum phase transitions into four-fold degenerate columnar VBS states. This phase transition can be investigated in
detail only with QMC simulations. SSE studies have established what seems like a continuous transition \cite{Sandvik07,Melko08,Sandvik10c,Kaul11,Block13,Pujari15},
similar to the proposed DQCP but with anomalous scaling corrections that so far only have a phenomenological explanation \cite{Shao16} . It has also been
proposed that the transition is actually weakly first-order, with the true DQCP only existing outside the space that can be reached with lattice models
(e.g., in a fractal dimension or on the complex plane) \cite{Wang17}. Though the ultimate nature of the phase transition is still unsettled, it is already
clear that the length scale associated with the transition is very large, and the DQCP phenomenology applies.

\vskip2mm
By perturbing the $J$-$Q$ model sufficiently so that the $Q$ terms form a checker-board pattern,
the VBS state can be modified from a four-fold columnar to a two-fold degenerate plaquette-singlet state. \index{plaquette-singlet state}
The transition from the antiferromagnet into the plaquette-singlet state is clearly first-order according to $T>0$ SSE and $T=0$ projector QMC
studies \cite{Zhao19}. However, the transition point is associated with an unexpected higher symmetry, combining the O(3) magnetic order parameter and
the scalar $Z_2$ plaquette order parameter into an O(4) vector. A similar phenomenon with emergent SO(5) symmetry has been studied with SSE simulations
in a spin-$1$ $J$-$Q$ model \cite{Wildeboer18}. The plaquette-singlet
state is of relevance in the frustrated 2D magnet SrCu$_2$(BO$_3$)$_2$ \cite{Zayed17}, and SSE simulations aimed at explaining the phase transitions driven
by high pressure in this system have been reported very recently \cite{Guo19}.

\vskip2mm
Building on the idea of the $J$-$Q$ model, Kaul and collaborators have constructed several other classes of sign-free ``designer Hamiltonians''
\cite{Kaul15a,Kaul15b,Demidio17,Desai19,Block19}. The original $J$-$Q$ models and these extended variants provide unique opportunities
to further explore many interesting quantum phases and quantum phase transitions.

\vskip2mm
Highly frustrated Ising interactions, supplemented with various off-diagonal terms, can also be studied with SSE simulations (though the sampling
is more challenging \cite{Melko07} and system sizes as large as those for Heisenberg and $J$-$Q$ interactions can not be reached). The focus of these studies is
typically to explore different incarnations of $Z_2$ quantum spin liquids and their quantum phase transitions \cite{Dang11,Isakov12,Becker18}.

\subsection{Dynamic response of quantum magnets}
\label{sec:dynamics}
           
To connect numerical simulations of lattice models to experiments, dynamic response functions are the most useful, e.g., in quantum magnets the
dynamic spin structure factor $S(q,\omega)$ can be measured directly in inelastic neutron scattering experiments, \index{dynamic structure factor}
and the low-energy structure factor
is accessed, e.g., in NMR experiments. In QMC calculations, dynamic spectral functions can only be accessed in the form of imaginary-time dependent
correlation functions, and these have to be analytically continued to the real time (or frequency) domain using some numerical scheme \cite{Jarrell96}.
Analytic continuation in its own right is an interesting and challenging technical
problem subject to ongoing research activities; see Ref.~\cite{Sandvik16,Shao17} for a recent example of methods
developed by the author and collaborators. While all numerical analytical continuation method have natural limitations in the frequency resolution
that can be achieved, due to the statistical noise in the QMC data (even when the noise level is exceedingly small), important spectral features
can be resolved, and some times it is possible to compare with experiments in a very detailed manner \cite{Shao17,Ying19}.

\vskip2mm
While the static properties of the 2D Heisenberg model have been well understood for some time, there has been a long-standing unresolved mystery
in the dynamic response: At and close to the equivalent wave-vectors $q=(\pi,0)$ and $(0,\pi)$, the excitation energy is reduced and the spectral line shape
of $S(q,\omega)$ is anomalously broadened. The anomalies cannot be easily explained within spin-wave theory. In recent work based on SSE and analytic continuation,
it was found that the phenomenon is a precursor to a DQCP that can be reached by adding other interactions \cite{Shao17}. Spectral functions
at the DQCP of the $J$-$Q$ model have also been studied and are in excellent agreement with a field-theory treatment based on the so-called
$\pi$-flux state \cite{Ma18b}.

\vskip2mm
The simpler dimerized Heisenberg models with O(3) transitions also have interesting dynamical response functions. In particular, the amplitude
model (also often called the Higgs mode) of near-critical systems (on the ordered side of the phase transition) have been studied in 2D \cite{Lohofer15}
and 3D \cite{Lohofer17,Qin17} and compared with experimental results for  TlCuCl$_3$ \cite{Merchant14}.

\vskip2mm
In highly frustrated quantum Ising systems, spectral functions can give important insights into the nature of exotic excitations. Recent SSE-studied
expamples include the identification of separate photon and spinon modes in a quantum spin-ice system \cite{Huang18} and a two-spinon continuum in a
kagome-lattice model \cite{Becker18}.

\subsection{Disordered systems}
\label{sec:disorder}

\index{quenched disorder}
Randomness (quenched disorder) can fundamentally affect quantum states and quantum phase transitions. Many SSE studies have been devoted to the
effects of random couplings in the ordered Heisenberg antiferromagnet \cite{Laflorencie06} and at the $O(3)$ transition in the dimerized Heisenberg
systems mentioned above in Sec.~\ref{sec:critical} \cite{Yu05,Sandvik06,Tan17}. A still open issue is why the Harris criterion for the relevance or
irrelevance of disorder appears to be violated in some cases \cite{Ma14}. Systems with dilution have also been frequently studied \cite{Sandvik02b,Sandvik02c}, and
interesting excitations with a very low energy scale, that was not anticipated, have been found at the percolation point on the 2D square lattice
\cite{Wang10}.

\vskip2mm
Very recently, effects of various types of disorder at the DQCP in the $J$-$Q$ model have been investigated \cite{Liu18}, and it was
found that the VBS state is replaced by a critical phase similar to the {\it random singlet state} that is well known in random $S=1/2$ Heisenberg chain
(and which have also been studied with SSE simulations \cite{Shu16,Shu18}). SSE simulations have also been applied to the Bose-Hubbard model with site
randomness \cite{Wang15b}, where interesting differences were found between weak and strong disorder.

\vskip2mm \index{quantum annealing}
Disordered frustrated Ising models, which often have spin-glass phases,
are of interest in the field of optimization with quantum annealing. Excitation gaps have been extracted from
imaginary-time correlations computed with the SSE method \cite{Farhi12}, and the generalized SSE method for imaginary-time quantum annealing has also been used to
study the dynamics of the quantum phase transition between a paramagnet and a quantum spin glass \cite{Liu15}. Recent further developments of the SSE
algorithm for highly anisotropic (near-classical) frustrated Ising models were specifically aimed at quantum annealing applications \cite{Albash17}. 
\vskip5mm

\subsection*{Acknowledgments}
Some of the work reported here was supported by the NSF under Grant No.~DMR-1710170, by a Simons Investigator Award, and by computer
time allocations on the Shared Computing Cluster administered by Boston University's Research Computing Services.



\clearchapter


\end{document}